\documentclass[slac_one]{revtex4}
\usepackage{graphicx,epsfig,tabularx}
\usepackage{fancyhdr}
\usepackage{amsmath,amsfonts}
\newcommand{\lesim}{\,\raisebox{-.3ex}{$_{\textstyle
<}\atop^{\textstyle\sim}$}\,}
\newcommand{\gesim}{\,\raisebox{-.3ex}{$_{\textstyle
>}\atop^{\textstyle\sim}$}\,}

\newcommand{\MW}{M_{\text{W}}}
\newcommand{\MZ}{M_{\text{Z}}}

\newcommand{\mt}{m_{\text{t}}}

\newcommand{\cha}{\tilde{\chi}}
\newcommand{\neu}{\tilde{\chi}^0}
\newcommand{\mcha}[1]{m_{\tilde{\chi}^\pm_{#1}}}
\newcommand{\mneu}[1]{m_{\tilde{\chi}^0_{#1}}}

\newcommand{\gev}{{\rm \ GeV}}
\newcommand{\tev}{{\rm \ TeV}}
\newcommand{\mev}{{\rm \ MeV}}

\begin{document}

\title{{\small{Hadron Collider Physics Symposium (HCP2008),
Galena, Illinois, USA}}\\ 
\vspace{12pt}
Status of Constraints on Supersymmetry} 

%

\author{A. Freitas}
\affiliation{Department of Physics \& Astronomy, University of Pittsburgh, PA
               15260, USA}

\begin{abstract}
A short summary of constraints on the parameter space of supersymmetric models
is given. Experimental limits from high energy colliders, electroweak
precision data, flavor and Higgs physics, and cosmology are considered. The main
focus is on the MSSM with conserved R- and CP-parity and minimal
flavor violation, but more general scenarios and extended models will also 
be discussed briefly.
\end{abstract}

\maketitle

\thispagestyle{fancy}


\section{INTRODUCTION}

The purpose of this contribution is to summarize the constraints on
supersymmetric models from various experimental results.
Due to the large wealth of experimental searches for physics beyond the standard
model (SM) and phenomenological studies on supersymmetry (SUSY)
it is impossible to cover all of them in this short review. Thus the author
apologizes that many valuable studies are not mentioned or cited in
this report.

To set the scene, a short review of the most widely studied SUSY models is given 
in the next section. The following sections discuss constraints on the
parameter space of these models from high energy colliders, electroweak
precision data, flavor and Higgs physics, and cosmology, respectively. Finally,
some qualitative comments on more general SUSY models are presented before the
summary.

\section{SUSY MODELS} 

The most extensively studied SUSY model is the Minimal Supersymmetric Standard
Model (MSSM), with the particle content listed in Table~\ref{mssm}.
In addition the MSSM imposes R-parity, assigning $R_p = +1$ for the Higgs boson,
gauge bosons, leptons, and quarks, and $R_p = -1$ for their supersymmetric
partners (neutralinos, charginos, gluino, sleptons, and squarks).
As a result the superpotential has the form
\begin{equation}
W_{\rm MSSM} = y_{\rm u} \hat{Q} \cdot \hat{H}_2 \, \hat{U}^c 
         + y_{\rm d} \hat{Q} \cdot \hat{H}_1 \, \hat{D}^c
         + y_{\rm e} \hat{L} \cdot \hat{H}_1 \, \hat{E}^c
	 - \mu \hat{H}_1 \cdot \hat{H}_2 \,.
\label{wmssm}	 
\end{equation}
For a general introduction to the MSSM and notational definitions, see {\it
e.$\,$g.} Ref.~\cite{martin}.

In the major part of this work, an even more minimal version of the MSSM is
assumed where the CKM matrix is the only source of CP violation and flavor
violation. In other words, the SUSY breaking parameters are assumed to be real
and flavor blind.
\begin{table}[t]
\begin{center}
\caption{Particle content of the MSSM}
\begin{tabular}{|@{$\quad$}c@{$\quad$}|@{$\quad$}c@{$\quad$}|@{$\quad$}c@{$\quad$}|}
\hline \textbf{Spin 0} & \textbf{Spin 1/2} & \textbf{Spin 1}
\\
\hline \hline
Neutral Higgses & Neutralinos & Photon $\gamma$ \\
 $h_0,\,H_0,\,A_0$ & $\neu_1 \dots \neu_4$ & $Z$ boson \\[.5ex]
\hline
Charged Higgs $H^\pm$ & Chargino $\cha^\pm_1, \cha^\pm_2$ & $W^\pm$ bosons \\[.5ex]
\hline
 & Gluino $\tilde{g}$ & gluon $g$ \\[.5ex]
\hline
\hline
sleptons $\tilde{e}$, $\tilde{\mu}$, $\tilde{\nu}$,... &
leptons $e$, $\mu$, $\nu$, ... &
\\
squarks $\tilde{u}$, $\tilde{d}$, ... &
quarks $u$, $d$, ... & \\
\hline
\end{tabular}
\label{mssm}
\end{center}
\end{table}

This still leaves more than one dozen {\it a priori} unknown SUSY breaking
parameters. Many experimental searches and phenomenological analyses thus
consider specific SUSY breaking scenarios:
\begin{itemize}
\item \emph{mSUGRA/CMSSM:} 
In \emph{minimal Supergravity} (mSUGRA) or the \emph{constrained MSSM} (CMSSM)
the scale of SUSY breaking is situated near the scale of gauge coupling
unification, $M_{\rm GUT} \approx 2\times 10^{16}\gev$. At this scale, there is one common
mass parameter each for the gauginos, scalars and triple-scalar couplings
($A$-terms), respectively. At lower energies, a more complex SUSY mass spectrum
emerges due to renormalization group running. As a result, the colored SUSY
partners (squarks and gluino) are substantially heavier than the weakly coupled
SUSY particles. The lightest SUSY particle (LSP) is typically the lightest
neutralino $\neu_1$, with $\mneu{1} \sim {\cal O}(100 \gev)$.
\item \emph{GMSB:}
In \emph{gauge mediated SUSY breaking} (GMSB)
the breaking of supersymmetry is transmitted by gauge interactions.
The minimal version, which introduces messengers in the fundamental
representation of SU(5), produces ${\cal O}(100 \gev)$ SUSY masses for a
messenger scale $\Lambda_{\rm mess} \sim 100 \tev$. Similar to mSUGRA, the gauge
couplings and gaugino masses unify at $M_{\rm GUT}$, but the sfermion masses do
not unify at any scale. The triple-scalar couplings ($A$-terms) are almost zero
at the messenger scale $\Lambda_{\rm mess} \sim 100 \tev$ and remain relatively
small at the electroweak scale. In GSMB, the LSP is typically the gravitino,
with $m_{\tilde{G}} \sim 100 {\rm \ eV} \dots 1 \gev$.
\item \emph{AMSB:}
In general, soft supersymmetry breaking terms receive contributions from the
super-Weyl anomaly via loop effects. 
\emph{Anomaly mediated
supersymmetry breaking} (AMSB) becomes relevant only if other SUSY breaking
mechanisms are suppressed or absent. AMSB
predicts the gaugino mass ratios $|M_1| : |M_2| : |M_3| \approx 2.8 : 1 : 7.1$,
so that the LSP 
is typically the lightest neutralino $\neu_1$ with a dominant wino
component. The chargino $\cha^\pm_1$
is a almost pure wino and very close in mass to the LSP.
\end{itemize} 

A shortcoming of the MSSM is the appearance of the $\mu$-term (the last term in
eq.~(\ref{wmssm})) which must be of the order of the electroweak scale for
successful electroweak symmetry breaking, leading to the unnatural hierarchy
$\mu \ll M_{\rm GUT}$. One solution to this puzzle is the introduction of an
additional singlet chiral superfield so that the general superpotential becomes
\begin{equation}
W_{\rm MSSM+S} = \lambda \hat{S} \hat{H}_1 \cdot \hat{H}_2 + \kappa \hat{S}^3 + 
  m_{\rm S} \hat{S}^2 +
  t_{\rm S} \hat{S} +
  \mbox{Yukawa terms}.
\label{smssm}	 
\end{equation}
In this general form the superpotential again has several dimensionful
parameters which have to be much smaller than the GUT scale. However, the 
unwanted terms can be set to zero by introducing new symmetries, for example
\begin{itemize}
\item \emph{Next-to-minimal MSSM (NMSSM):}
 A global $\mathbb{Z}_3$ symmetry mandates $m_{\rm S}=t_{\rm S}=0$, but could
 lead to cosmological domain walls \cite{domain}.
\item \emph{Nearly minimal MSSM (nMSSM):}
 Imposing a global  $\mathbb{Z}_5$ or $\mathbb{Z}_7$ symmetry forbids all
 singlet self-couplings at tree-level, $m_{\rm S}=t_{\rm S}=\kappa=0$. 
 However, supergravity effects combined with SUSY breaking allow a contribution
 to $t_{\rm S}$ at the six- or seven-loop level, naturally generating a value
 $t_{\rm S} \sim {\cal O}$(TeV) as required for successful electroweak symmetry
 breaking \cite{nMSSM}.
\item \emph{U(1)-extended MSSM (UMSSM):} 
 This model introduces a U(1) gauge symmetry under which the Higgs and singlet field are
 charged. As a result, $m_{\rm S}=t_{\rm S}=\kappa=0$, but new D-term
 contributions to the Higgs potential appear which play an important role in
 achieving realistic electroweak symmetry
 breaking \cite{UMSSM}.
\end{itemize}

\section{HIGH ENERGY COLLIDERS}

Searches for SUSY particles at $e^+e^-$ colliders are largely independent on the
details of the model or scenario. Roughly speaking, results from LEP exclude
sparticles up to the beam energy $E_{\rm beam} \sim 100 \gev$. The actual
exclusion bounds \cite{lepsusy,pdg} are listed in Table~\ref{lepsusy}.
The exact limits vary as a result of the different pair-production cross
sections for different particles types.
Furthermore, some of the searches fail if the mass difference between the
pair-produced sparticle $\tilde{X}$ and the LSP becomes too small,
$m_{\tilde{X}} - m_{\rm LSP} \lesim$ few GeV, see Refs.~\cite{lepsusy,pdg} for 
details.
\begin{table}[t]
\begin{center}
\caption{Lower limits on SUSY particle masses from LEP searches}
\begin{tabular}{|@{$\;\;$}l@{$\quad$}rr|@{$\;\;$}l@{$\quad$}rr|}
\hline \textbf{Sparticle} & \textbf{lower limit} & \textbf{[GeV]} & 
\textbf{Sparticle} & \textbf{lower limit} & \textbf{[GeV]}
\\
\hline
$\;\;\neu_2$ & 62.4 && $\;\;\tilde{\nu}$ & 94.0 & \\
$\;\;\neu_3$ & 99.9 && $\;\;\tilde{e}_L$ & 107.0 & \\
$\;\;\neu_4$ & 116.0 && $\;\;\tilde{\mu}_R$ & 91.0 & \\
$\;\;\cha^\pm_1$ & 94.0 && $\;\;\tilde{\tau}_1$ & 81.9 & \\
$\;\;\tilde{u},\tilde{d},\tilde{c},\tilde{s}$ & 97.0 && 
	$\;\;\tilde{t}_1$ & 92.6 & \\
\hline
\end{tabular}
\label{lepsusy}
\end{center}
\end{table}

SUSY searches at hadron colliders are more intricate due to the large
backgrounds. In most cases, a large signal-to-background ratio is only
achievable by designing the selection strategy for some set of SUSY scenarios.
For SUSY searches at the Tevatron, mSUGRA/CMSSM scenarios are usually taken as
benchmark \cite{tevsusy}. However, when these results are expressed for more
general MSSM scenarios the limits become much weaker and might drop below the
LEP limits
\cite{tevsusy,alwall}, see
Table~\ref{tevlim}. More details about Tevatron searches and prospects for SUSY
discovery at the LHC are given in the contributions by T.~Adams \cite{adams} and
O.~Brandt \cite{brandt} at this conference.
\begin{table}[t]
\begin{center}
\caption{Lower limits on SUSY particle masses from Tevatron searches}
\begin{tabular}{|@{$\;\;$}c@{$\quad$}c@{$\quad$}c|}
\hline \textbf{Sparticle} & \textbf{mSUGRA/CMSSM} & 
\textbf{more general MSSM} \\
 & \textbf{limit [GeV]} & 
\textbf{limit [GeV]}
\\
\hline
$\tilde{g}$ & 308 & $\sim 150$ \\
$\tilde{q}$ & 380 & LEP limit \\
$\cha^\pm_1$ & 140 & LEP limit \\
\hline
\end{tabular}
\label{tevlim}
\end{center}
\end{table}

\vspace{1em}
The MSSM parameter space is also constrained indirectly by the lower limit on 
the mass of a SM-like Higgs boson from LEP searches, $m_{\rm h}^{\rm SM} >
114.4\gev$ \cite{lephiggs}.
In the MSSM the lightest CP-even Higgs boson mass can be calculated as a
function of other parameters. The leading tree-level and one-loop contributions
are given by
\begin{equation}
m_{\rm h} \lesim \MZ^2 \, \cos^2 2\beta + \frac{3\mt^4}{2\pi^2v^2}
\left[ \log \frac{m_{\tilde{t}}^2}{\mt^2} + \frac{X_t^2}{m_{\tilde{t}}^2}
\right]
+ ... ,
\qquad
X_t = A_t - \frac{\mu}{\tan\beta},
\label{hfor}
\end{equation}
where the dots stand for higher-order corrections. To be compatible with the LEP
limit, the terms in eq.~(\ref{hfor}) need to be large so that
at least one of the following conditions must be met:
\begin{itemize}
\item $\tan\beta \gg 1$ to maximize the tree-level term $\MZ^2 \, \cos^2
2\beta$,
\item Large average stop mass, $m_{\tilde{t}}^2 = m_{\tilde{t}_1} m_{\tilde{t}_2}
\gesim (1 \tev)^2$,
\item Large stop mixing to enhance the $X_t^2/m_{\tilde{t}}^2$ term.
\end{itemize}

In extended models with extra singlets (NMSSM, nMSSM) or gauge groups (UMSSM)
the constraints on the SUSY parameter space from the $m_{\rm h}$ limit are much
less severe due to new positive tree-level contributions to $m_{\rm h}$
\cite{Batra:2004vc}.

\section{ELECTROWEAK PRECISION DATA}

Loop corrections from SUSY particles affect the predictions for electroweak
precision observables. Much effort has been invested in calculating the
radiative corrections from SUSY loops. The current state of the art in the MSSM
encompasses complete one-loop corrections \cite{ewal} and leading two-loop
corrections of order ${\cal O}(\alpha \alpha_s)$ \cite{ewalals}
and ${\cal O}(\alpha y_{t,b}^2)$ \cite{ewaly}. For the NMSSM and other
extensions only partial one-loop results are known.

The most important quantities that have been measured with high precision and
that receive sizable corrections from new physics are:
\begin{itemize}
\item The $W$-boson mass $\MW$, which is determined from the muon decay width by
including the relevant radiative corrections for the process 
$\mu^- \to e^- \bar{\nu}_e \nu_\mu$.
\item The effective weak mixing angle of the $Z$ boson, defined through the
effective vector and axial vector couplings of the $Z$ boson on the $Z$
resonance, $\sin \theta_{\rm eff} =
\frac{1}{4}\left(1-{\rm Re} \frac{v_{\rm eff}}{a_{\rm eff}} \right)$.
The effective mixing angle can be defined for all fermion flavors although the
numerical differences are small except for the $Zb\bar{b}$ vertex.
\item The total $Z$-boson width, $\Gamma_{\rm Z}$, and the total $Z$ peak cross
section $\sigma[e^+e^- \to Z \to f\bar{f}]$. Both of these quantities are
closely related to the coupling combination $v_{\rm eff}^2+a_{\rm eff}^2$.
\item The muon anomalous magnetic moment $a_\mu = (g_\mu-2)/2$.
\end{itemize}
By performing a fit of the MSSM predictions for these quantities to the
experimentally determined values \cite{weber,su} one finds in general good agreement for
light sleptons and gauginos. The reason for this is two-fold: {\it (i)} in the
SM the best fit to the electroweak precision observables corresponds to a Higgs mass of
$m_{\rm h} \approx 87\gev$, which creates a tension with the lower limit from
direct searches, $m_{\rm h} > 114.4\gev$. The new contributions from
slepton-gaugino loops can push the best-fit Higgs mass to values $m_{\rm h} >
100\gev$, thus improving the overall goodness-of-fit. {\it (ii)} SUSY loop
contributions from sleptons and gauginos can account for the 3.3$\sigma$
discrepancy between the SM prediction and the measured value of the muon anomalous
magnetic moment, $a_\mu^{\rm exp} - a_\mu^{\rm theo} = (27.5 \pm 8.4) \times
10^{-10}$ \cite{amu}.

\begin{figure}[t]
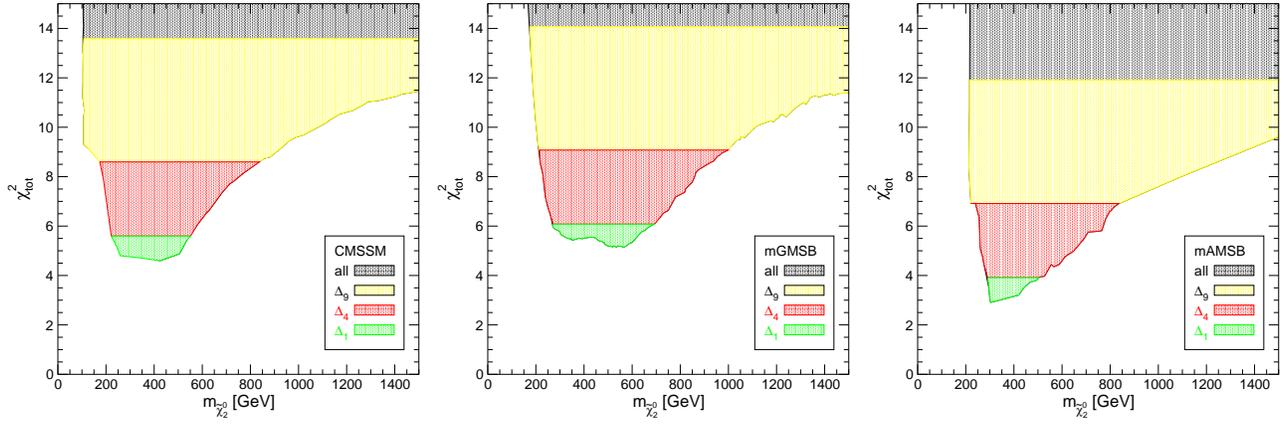

\centering
\epsfig{figure=asbs3_s_mass12_cl.eps, width=5.5cm} \ 
\epsfig{figure=asbs3_g_mass12_cl.eps, width=5.5cm} \ 
\epsfig{figure=asbs3_a_mass12_cl.eps, width=5.5cm}
\vspace{-2ex}
\caption{$\chi^2$ fit to electroweak precision observables as a function of the mass
of the second neutralino $\neu_2$ for the  mSUGRA/CMSSM, GMSB, and AMSB
scenarios; from Ref.~\cite{su}.} \label{ewsusy}
\end{figure}

The results of a $\chi^2$ fit for the mSUGRA/CMSSM, GMSB, and AMSB scenarios are
shown in Fig.~\ref{ewsusy}. The plots show the best fit $\chi^2$ as a function
of the mass of the neutralino $\neu_2$, which has a dominant wino or higgsino
component  in these scenarios.
As evident from the figure, in all three scenarios a light neutralino with
$\mneu{2} \sim 200 \dots 700 \gev$ is preferred, while neutralino masses above
1~TeV are strongly disfavored.

\section{FLAVOR AND HIGGS PHYSICS}

Rare decays of heavy flavor mesons are very sensitive to new physics effects.
For SUSY models these effects are enhanced for large values of $\tan\beta$ since
the Yukawa couplings of the down-type fermions become large, $y_{\rm d} = 
m_{\rm d}/v \times \tan\beta$. Schematically, the dependence of rare $B$ decays
on $\tan\beta$ in the MSSM reads
\begin{align}
{\rm BR}[B_s \to \mu\mu] &\sim \frac{\tan^6\beta}{M_A^4}, \\
{\rm BR}[B_u \to \tau\nu] &\sim \biggl [ 1-\frac{m_B^2}{M_A^2} \tan^2\beta \biggr ]^2, \\
{\rm BR}[b \to s\gamma] &\sim 1+ A\tan\beta+B\tan\beta/M_A^2,
\end{align}
where $A$ and $B$ are coefficient that depend on other SUSY parameters in a
non-trivial way.

In the region of large $\tan\beta$, the production cross section for the CP-odd
Higgs boson $A_0$ at hadron colliders is also increased,
\begin{equation}
\sigma[pp \to A \to \tau\tau] \sim \tan^2\beta,
\end{equation}
establishing an intricate relationship between heavy-flavor and Higgs
observables with respect to the SUSY parameter space \cite{menon}.

The negative searches of the $A_0$ boson at the Tevatron \cite{a0cdf,a0d0}
exclude the parameter
region of large $\tan\beta$ and small $M_A$, with very little dependence on
other SUSY parameters. Similarly, the current experimental
upper limit on BR$[B_s \to \mu\mu]$ is most important for 
large $\tan\beta$ and  small $M_A$,
although with some dependence on other variables, $m_{\tilde{t}_{1,2}},
m_{\tilde{b}_{1,2}}, \mu$. Depending on the value of these quantities
the constraint from BR$[B_s \to \mu\mu]$ on the SUSY parameter space can be
stronger or weaker than the bound from  $A_0$ searches.
The measurement of BR$[B_u \to \tau\nu]$, in good agreement with the SM, allows
either the region of small $\tan\beta/M_A$ or of large $\tan\beta/M_A$ (although
the latter is severely limited by the previous two observables), while
intermediate values of $\tan\beta/M_A \sim 1/4 \gev^{-1}$ are disfavored.
Finally, BR$[b \to s\gamma]$ varies relatively mildly as a function of 
$M_A$, but it places both an lower and upper bound on $\tan\beta$.
However, the prediction of  BR$[b \to s\gamma]$ is affected by many SUSY parameters
so that quantitative conclusions depend quite strongly on the scenario.

\begin{figure}[t]
\centering
\epsfig{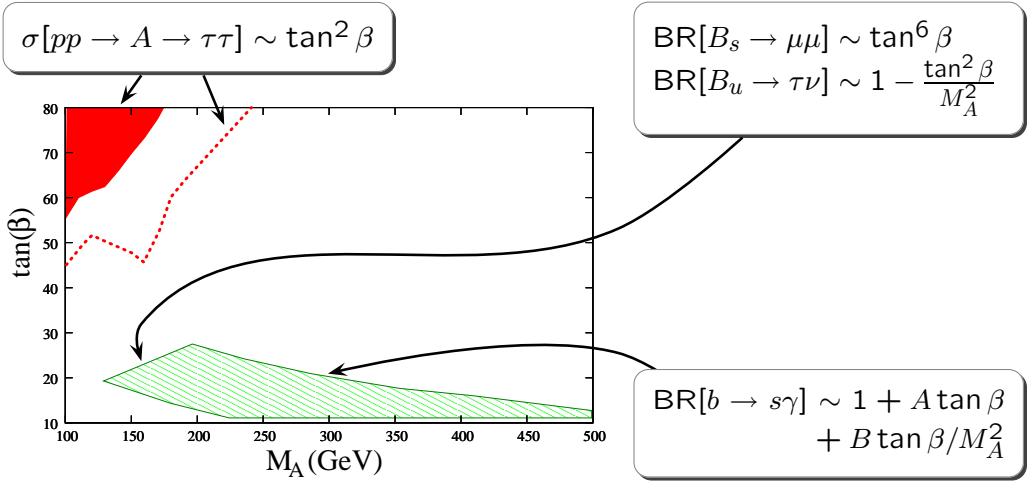}
\vspace{-2ex}
\caption{Constraints on the $M_A$--$\tan\beta$ parameter plane from $B$ and
Higgs physics in the MSSM; from Ref.~\cite{menon}. The other SUSY parameters are chosen
to be $\mu = -100\gev$, $X_t = 2.4\tev$, $m_{\tilde{q}}=1\tev$,
$m_{\tilde{g}}=800\gev$.
The red region is excluded by non-observation of $A_0$ production
at the Tevatron (solid filled: CDF \cite{a0cdf}; dotted: D$\emptyset$ \cite{a0d0}), while only 
the green region is allowed by rare $B$-decay measurements.} \label{bh}
\end{figure}

The flavor physics and Higgs constraints mentioned above are summarized in
Fig.~\ref{bh} for the MSSM.
Note that while the $B$-physics observables
seem to impose very severe limits on $M_A$ and
$\tan\beta$
these bounds depend substantially on other SUSY parameters $\mu$, 
$m_{\tilde{g}}$, $X_t$, and $m_{\tilde{q}}$, which for the purpose of this
analysis is assumed to be a common mass for all squarks.
The most robust, scenario-independent constraint comes from $A_0$ searches
which can be expressed roughly as $M_A/\tan\beta \gesim 3\gev$.

\section{COSMOLOGY}

The derivation of bounds on the SUSY parameter space from cosmology depends on
many details of the SUSY model as well as the history of the universe and might
be impacted by theoretical uncertainties that have not been quantified so far.
Nevertheless it is illustrative to study some of the constraints since even at a
qualitative level they affect the parameter structure of the model.

\subsection{Dark Matter}

For conserved R-parity, the LSP is a stable particle and could provide a good
cold dark matter candidate as long as it is neutral and weakly interacting.
Within the standard cosmological model it is possible to calculate the expected
relic dark matter density for a given SUSY model, although often the results
depend on many model parameters. However typically only certain corners of the
parameter space give good agreement with the measured value from the cosmic
microwave background, $\Omega_{\rm DM} h^2 = 0.110 \pm 0.006$ \cite{wmap}.
There are three main possibilities for LSPs as 
viable dark matter candidates in the MSSM:
\begin{itemize}
 \item \emph{Lightest neutralino $\neu_1$:}
 If the lightest neutralino has a dominant bino component, 
 annihilation into gauge bosons is strongly suppressed. 
 Thus, to be compatible with
 the observed dark matter density, one of the following enhancement mechanisms for
 the annihilation cross section needs to be present:
 \begin{itemize}
  \item Light sleptons, $m_{\tilde{l}} \approx 100\gev$, together with
  	$\mneu{1} < 100\gev$ lead to a sufficiently large t-channel
	contribution. This parameter region is often called the \emph{``bulk''
	region}.
  \item Co-annihilation: if the mass difference to the next-to-lightest SUSY
  particle (NSLP) $\tilde{X}$ is small, $m_{\tilde{X}}-\mneu{1} \ll \mneu{1}$,
  both particles annihilate in parallel in the early universe. A large
  $\tilde{X}\neu_1$ co-annihilation cross section can then compensate for a small
  $\neu_1\neu_1$ annihilation rate.
  \item Resonant annihilation: if the mass of the neutralino is close to half of
  the mass of a possible bosonic s-channel resonance, $2\mneu{1} \approx \MZ, \,
  m_{\rm h}, \, M_A$,  neutralino pair annihilation can proceed efficiently
  through this resonance.
 \end{itemize}
 Alternatively, the neutralino $\neu_1$ could be an admixture with sizable wino
 and/or higgsino components. In this case, the $\neu_1\neu_1$ annihilation rate
 into gauge boson naturally has the right order of magnitude for $\mneu{1} \sim
 {\cal O}({\rm few\ 100\ GeV})$.
 \item \emph{Sneutrino $\tilde{\nu}$:}
 It has been known for many years that the L-sneutrino $\tilde{\nu}_L$ cannot be
 the dominant source of dark matter since it would lead to a collision rate
 with ordinary matter that is much larger than the current bounds from direct detection
 experiments.
 However, as indicated by the observation of neutrino oscillations, 
 it is likely that also L- and R-sneutrinos mix with each other. A sneutrino
 with a dominant R-sneutrino ($\tilde{\nu}_R$) component would constitute a good
 dark matter candidate in agreement with all constraints for 10 GeV $\lesim
 m_{\tilde{\nu}_R} \lesim$ 1 TeV \cite{sneu}.
 \item \emph{Gravitino $\tilde{G}$:}
 Gravitino dark matter can be produced in two ways, see {\it e.$\,$g.}
 \cite{steffen}:
 \begin{itemize}
  \item Gravitinos can be produced non-thermally from decays
  of the NLSP $\tilde{X}$. Late decays of the NLSP can lead to entropy overproduction and
  thus hot dark matter in disagreement with large scale structure formation.
  Since 
  \begin{equation}
  \Gamma[\tilde{X} \to X \tilde{G}] \propto
   \frac{m_{\tilde{X}}^5}{m_{\tilde{G}}^2} 
   \biggl(1-\frac{m_{\tilde{G}}^2}{m_{\tilde{X}}^2}\biggr)^4
  \end{equation} 
  this places
  a lower bound $m_{\tilde{X}} > 0.5\tev$. If this bound is satisfied the
  correct relic abundance can be obtained for gravitino masses in the range
  $1\gev \lesim m_{\tilde{G}} \lesim 700 \gev$ \cite{steffen}.
  \item Alternatively, gravitinos can be produced thermally directly from the
  hot plasma in the early universe. When produced from thermal
  equilibrium the gravitino abundance is much too large (``gravitino problem'').
  Therefore the reheating temperature $T_{\rm R}$ of the universe is required to be much
  smaller than the gravitino equilibrium temperature.
  In this case non-equilibrium thermal production is viable for $1{\rm \
  keV} \lesim m_{\tilde{G}} \lesim 1 \tev$, depending on the exact value of $T_{\rm R}$.
 \end{itemize}
\end{itemize}

\subsection{Big-bang Nucleosynthesis}

If the LSP is a gravitino the energy released from NLSP decays can be
problematic for successful big-bang nucleo\-synthesis (BBN). Hadronic and
electromagnetic showers emitted by the NLSP decays can dissociate light element
nuclei and thus shift the predicted ratios of element abundances. For the NLSP
to disrupt BBN, the NLSP lifetime has to be $\tau_{\rm NLSP} \gesim 100$~s.
Therefore this constraint excludes small NLSP masses and large gravitino masses.

\begin{figure}[t]
\centering
\epsfig{figure=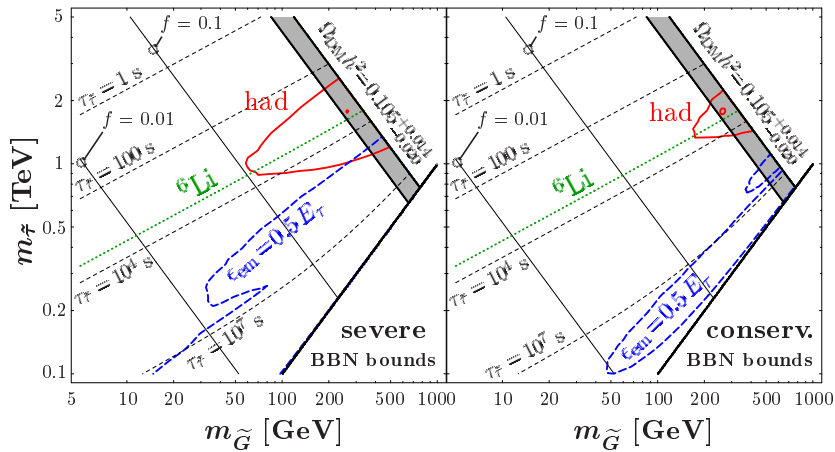, width=11cm}
\vspace{-2ex}
\caption{Constraints on the $\tilde{\tau}_1$ and $\tilde{G}$ masses from
BBN due to hadronic and electromagnetic energy release from late
$\tilde{\tau}_1$ decays; from Ref.~\cite{steffen2}. The dashed lines show
contours of equal $\tilde{\tau}_1$ stau lifetime, while the gray band indicates
the region compatible with the correct relic density. The difference between the
left and right plot serve as an illustration of the uncertainty for primordial
nuclei abundances.} \label{bbn}
\end{figure}

For $\tilde{\tau}_1$ as NLSP, the detailed constraints are shown in
Fig.~\ref{bbn} \cite{steffen2}.
Depending on assumptions in the evaluation of the primordial nuclei abundances
the BBN constraints place an upper bound $m_{\tilde{G}} \lesim 100$--500 GeV.

\subsection{Baryogenesis}

New physics beyond the SM is needed to explain the excess of matter over
antimatter in the universe. The two most well-known mechanisms are leptogenesis
and electroweak baryogenesis. \emph{Leptogenesis} generates the
particle-antiparticle asymmetry through the decay of long-lived heavy neutrinos
$\nu_R$ or sneutrinos $\tilde{\nu}_R$. This mechanisms imposes strong
constraints on the masses, mixings and CP phases of the right-chiral (s)neutrino
sector, but if $m_{\nu_R} \gg 1\tev$ it is in general not testable by collider
experiments. 
In \emph{electroweak baryogenesis}, on the other hand, the
matter asymmetry is created by the electroweak phase transition if it is
strongly first order and involves CP-violating currents. 

In the MSSM a strong first order phase transition is only realizable if one of
the stops is light, $m_{\tilde{t}_1} < 140\gev$ \cite{mssmbg}. The Higgs mass
bound, $m_{\rm h} > 114.4 \gev$ then requires the other stop to be much heavier,
$m_{\tilde{t}_2} > 3\tev$. In singlet extensions (NMSSM/nMSSM) the
strength of the electroweak phase transition is increased by the new
Higgs-singlet couplings and no special values for the stop masses are needed
\cite{nmssmbg}.

CP-violating currents can originate from the chargino/neutralino sector both in
the MSSM and NMSSM/nMSSM. However, due to strong limits on electric dipole
moments of the electron and neutrino, such CP phases are only allowed for very
large masses of the first generation sfermions, $m_{\tilde{e}}, m_{\tilde{q}}
\gesim 10\tev$. In the NMSSM/nMSSM the CP phase responsible for baryogenesis can
also be implemented in the Higgs sector, leading to weaker constraints from
electric dipole moments \cite{huber}.

\subsection{Ultra-light neutralinos}

Neutralinos $\neu_1$ that are almost exclusively bino and have negligible wino
and higgsino components are not constrained by collider data. The only relevant
bounds come from astrophysics and cosmology \cite{dreiner}:
\begin{itemize}
\item For conserved R-parity a lower bound $\mneu{1} > 3\gev$ has to be
imposed to avoid dark matter overproduction.
\item Independent of R-parity conservation, very light neutralinos can contribute
to supernova cooling for moderately light selectrons, $m_{\tilde{e}} \lesim
500\gev$. The observations from SN~1987A thus lead to a lower limit $\mneu{1} >
200\mev$ in this case. However, for $m_{\tilde{e}} > 1200\gev$ no constraint on
the neutralino mass can be derived from supernova cooling.
\item Very light neutralinos have a large free-streaming length and thus can
jeopardize structure formation. This consideration excludes values of $\mneu{1}$
between 1~eV and 1~keV.
\end{itemize}
In summary, limits on light bino-like $\neu_1$ are very weak and $\mneu{1}$ is
largely unconstrained.

\section{EXTENDED MODELS}

In this section the assumptions of R-parity conservation,  minimal flavor
violation and CP conservation will be relaxed one at the time. As a result, many
bounds on the MSSM parameter space become weaker or disappear altogether.

\subsection{Flavor violation}

The sfermion soft breaking parameters can introduce new sources of flavor
violation, in particular leading to potentially large flavor changing neutral
currents (FCNCs). FCNCs are strongly constrained by $K^0$, $D^0$ and $B^0$
mixing, rare $B$ decays, and limits on lepton flavor violating processes such as
$\mu \to e\gamma$, $\mu \to e$ conversion, {\it etc.}
However, if new flavor violating terms are introduced in the 2nd and 3rd
generation only, flavor mixing sfermion mass terms as large as ${\cal O}(M_{\rm
SUSY})$
are still allowed by present data \cite{utfit}.

\subsection{CP violation}

Complex CP phases in the gaugino sector and in the parameters of the 1st
generation sfermions are strongly constrained by electric dipole moments (see
previous section). Sizable CP violation is however allowed in the Higgs sector
and the sector of the 3rd generation sfermions.

\subsection{R-parity violation}

Without R-parity conservation the MSSM superpotential is extended by the
following couplings:
\begin{equation}
W_{{\not{R_p}}\rm MSSM} = W_{\rm MSSM}
+ \tfrac{1}{2} \lambda_{ijk} L_i \cdot L_j E_k^c
+ \tfrac{1}{2} \lambda'_{ijk} L_i \cdot Q_j D_k^c
+ \tfrac{1}{2} \lambda''_{ijk} U_i^c \cdot D_j^c D_k^c.
\end{equation}
The product of baryon-number violating and lepton-number violating couplings is
strongly constrained by proton decay, {\it i.$\,$e.}
$|\lambda''_{ijk}\lambda_{ijk}|,
|\lambda''_{ijk}\lambda'_{ijk}| \ll 1$. Such a structure could be explained by
discrete symmetries while still allowing some non-zero R-parity violating
couplings. In the absence of $B$-violating terms the $L$-violating interactions
are mainly constrained by data on neutrino masses, leading to
$|\lambda_{ijk}|,|\lambda'_{ijk}| \lesim
10^{-5} \dots 0.6$ \cite{rpl}.

If R-parity is violated the LSP is not stable and the signatures for SUSY
particles production at colliders are dramatically altered. As a result,
experimental bounds for several SUSY particles becomes much weaker. In
particular one finds $m_{\tilde{g}} > 51$ GeV \cite{schwartz}, $m_{\tilde{b}_1}
> 7.5$ GeV, $m_{\tilde{\tau}_1} > 11$ GeV \cite{janot}, and  $m_{\rm h} > 82$
GeV \cite{carpenter}.

\section{SUMMARY}

Due to the complexity of the SUSY parameters space (even in the MSSM with
R-parity conservation, minimal flavor violation and CP conservation) and the
large number of experimental results it is difficult to summarize all
constraints in a simple picture. 
In Table~\ref{summ} a rough overview of the main limits from direct sparticle
and Higgs searches, as well as electroweak precision data and flavor physics is
attempted. The three columns in the table correspond to the MSSM with and without
R-parity conservation and the NMSSM, respectively. The NMSSM limits also apply
for the nMSSM and UMSSM. A ``---'' indicates that no bound exists while ``?''
stands for cases where the final quantitative conclusion is not known yet.
\begin{table}[t]
\begin{center}
\caption{Simplified summary of parameter constraints in the MSSM without and
with R-parity violation and in the NMSSM.}
\begin{tabular}{|c|c|c|}
\hline
\makebox[4cm]{MSSM + $R_p$} & 
\makebox[4cm]{MSSM + ${\not{\!\!R_p}}$} & 
\makebox[4cm]{NMSSM \& other ext.} \\
\hline
\hline
\multicolumn{3}{|c|}{$m_{\tilde{l}}, \, m_{\tilde{q}}, \, \mcha{1} \gesim 100$
GeV,
$m_{H^\pm} \gesim 100$ GeV} \\
\hline
$m_{\tilde{g}} \gesim 150$ GeV &
$m_{\tilde{g}} \gesim 51$ GeV &
$m_{\tilde{g}} \gesim 150$ GeV \\
\hline
$m_{\tilde{\tau}_1} \gesim 82$ GeV &
$m_{\tilde{\tau}_1} \gesim 11$ GeV &
$m_{\tilde{\tau}_1} \gesim 82$ GeV \\
\hline
\multicolumn{3}{|c|}{$m_{\tilde{\mu}}, \, \mneu{2} < 1$ TeV} \\
\hline
$\mneu{1} > 3$ GeV & --- & ? \\
\hline
$\tan\beta \gesim 3$ & --- & --- \\
\hline
$\sqrt{m_{\tilde{t}_1}m_{\tilde{t}_2}} \gesim 1$ TeV & 
\multicolumn{2}{c|}{\raisebox{-1.5ex}{?}} \\
and/or large $A_t$ & \multicolumn{2}{c|}{} \\
\hline
\multicolumn{2}{|c|}{$M_A/ \tan\beta \gesim 3$ GeV} & --- \\
\hline
\end{tabular}
\label{summ}
\end{center}
\end{table}
The limits in the first three lines stem for direct searches at high-energy
colliders, while the upper bound in the fourth line originates from electroweak
precision data. The fifth line gives limits from astrophysics and cosmology on
light neutralinos. Other cosmological constraints are not included in the table.
Finally, the last three lines summarize bounds from
flavor and Higgs physics.



\end{document}